\documentclass[12pt]{article}
\usepackage{amssymb}

\begin{document}

\author{\textit{R. Vilela Mendes} \\
{\small Grupo de F\'{\i }sica-Matem\'{a}tica}\\
{\small \ Complexo Interdisciplinar, Universidade de Lisboa}\\
{\small \ Av. Gama Pinto, 2, 1699 Lisboa Codex, Portugal}\\
{\small \ e-mail: vilela@alf4.cii.fc.ul.pt}}
\title{On some dynamical problems related to boundary layer turbulence}
\date{}
\maketitle

\begin{abstract}
A number of simplified dynamical problems is studied in an attempt to
clarify some of the mechanisms leading to turbulence and the existing
proposals to control this transition.

A simplified set of boundary layer equations displays a solution that
corresponds to the rolls and streaks instability and exhibits its streamwise
localized nature. The effect of random phases as a device to control the
transfer of energy to the small scales is studied both for the shell and the
boundary layer models. In spite of the simplified nature of the models, they
also provide some insight on the prospects for active turbulence control by
external body forces.
\end{abstract}

\section{Introduction}

A realistic model for the transition to turbulence and for the mechanisms of
turbulent energy dissipation, requires not only an accurate control of the
fluid equations but also a consideration of the effect of different boundary
conditions and of the external forces. Nevertheless, as in other fields of
physics, a step towards the understanding and control of the phenomena is
the identification of the universal mechanisms which are always present,
independently of the particular system or the boundary conditions. One such
mechanism is associated to the role of coherent structures in the generation
of turbulent energy.

Flow visualization techniques\cite{Kline} \cite{Kim} \cite{Choi}, as well as
two-point correlation measurements\cite{Sirovich1}, have revealed that a
large amount of the near wall turbulent energy is associated to coherent
structures (streamwise rolls and streaks) and to their breakdown (bursts).
Of particular importance for the shear stress peaks at the wall, that are
associated to the bursting events, is the transfer of fast moving fluid to
the near wall region (sweeps) and the transfer of energy to small turbulence
scales at a rate faster than the normal Kolmogoroff cascade. Understanding
and modeling these phenomena is of practical importance because, stabilizing
or delaying the evolution of the coherent structures, might decrease the
energy dissipation rate. For this purpose, in Sect. 2, a set of simplified
boundary layer equations is developed, which provides a good degree of
analytical control over the coherent structure solutions. From the
approximate equations and their solutions, the generation and lifting effect
of the rolls-plus-streaks instability is clearly understood as well as its
streamwise localized nature.

Another mechanism of critical importance, in the development of turbulence,
is the transfer of energy from the large to the small scales. The rate of
transfer may be also be affected by the time scale of destabilization of the
coherent structures. The transfer of energy between different length scales
is favored by the establishment of precise phase relations between the
dynamical evolution of the intervening scales. This led several authors\cite
{Handler} \cite{Sirovich2} to suggest that randomizing the phases might have
a strong effect either on blocking the cascade of energy transfer or on
destabilizing some structures. In Sect.3 the effect of random phases is
studied, first in the shell model and then in the boundary layer equations.
The study of these two cases puts into evidence the reasons and conditions
for effectiveness of random phase control techniques. In the shell model,
phase randomization of almost any mode, has an effective blocking effect on
the energy transfer to the small scales. However, in the boundary layer
equations, the effect is not so striking and is only observed if the phases
that are randomized are those of some particular modes. The reason for this
different behavior is the local (in momentum space) nature of the
interactions in the shell model as opposed to the long-range mode
interactions in the boundary layer equations. The conclusion is that for
boundary layer turbulence control, phase randomization is effective only if
a fine tuning of the phase-randomized modes is done. When the phase
randomization is implemented by boundary conditions (artificial roughness,
for example) the need for a fine tuning of the randomized modes may hinder
the applicability of the method.

Finally, in Sect.4, the prospects for active control of the coherent
structures by external body forces is considered. Here again, the fact that
simplified boundary layer equations are used, provides a clear view of the
relevant dynamical mechanisms. Feasible control signals are derivatives at
the wall and external forces cannot be designed with complex wall-normal
dependencies. Taking these two facts into account the conclusion is that
control by external body forces has almost no effect on the development,
growth and break-up of the coherent structures. Therefore the turbulence
production rate is not significantly changed. However, by controlling a
small layer of non-active fluid near the wall, a reduction in skin friction
drag might be obtained.

\section{Wall region coherent structures from the boundary layer equations}

Numerical simulations\cite{Hamilton} and a detailed analysis of the flow
equations\cite{Waleffe} have uncovered a possible self-sustained mechanism
for the coherent structures in the wall region. Let $x$, $y$ and $z$ be the
streamwise, the wall normal and the spanwise coordinates. The process starts
from counter-rotating streamwise rolls which would decouple from the
streamwise velocity component if the flow were independent of $x$. However
the rolls redistribute the momentum, creating fluctuations in the streamwise
velocity (streaks). The spanwise inflections create an instability of the
streaks and this in turn leads to a nonlinear energizing of the original
streamwise rolls. It is probable that this self-sustaining scheme, discussed
by Waleffe\cite{Waleffe} for a sinusoidal wall-bounded shear flow, captures
the correct physical process. However, because of its fragmented
mathematical description, it is difficult to use it as a basis for
quantitative predictions and, for example, for the evaluation of effective
control mechanisms. It would be better to obtain a single solution of the
flow equations comprising all the effects (rolls, streaks and instability
growth). For the full Navier-Stokes equations such solution does not seem
easy to obtain. However, for the simplified boundary layer equations, the
problem is not so hard and the mechanism may be made more explicit. In
particular it will be seen why it is essential to take the $x-$dependence
into account from the start. The starting point is the Navier-Stokes
equation for an incompressible fluid 
\begin{equation}
\frac{\partial \widetilde{U}}{\partial t}+(\widetilde{U}.\nabla )\widetilde{U%
}=-\frac{1}{\widetilde{\rho }_{m}}\nabla \widetilde{p}+\widetilde{\nu }%
\triangle \widetilde{U}+F  \label{2.1}
\end{equation}
where $F$ is the body force which, in the case of an ionized fluid\cite
{Vilela1}, for example, is 
\[
F=\frac{\widetilde{\sigma }}{\widetilde{\rho }_{m}}\widetilde{E}+\frac{%
\widetilde{\sigma }}{c\widetilde{\rho }_{m}}\widetilde{U}\times \widetilde{B}
\]
with the continuity equation 
\begin{equation}
\frac{\partial \widetilde{\rho }_{m}}{\partial t}+\nabla \cdot \left( 
\widetilde{\rho }_{m}\widetilde{U}\right) =0  \label{2.2}
\end{equation}
$\widetilde{\rho }_{m}$ being the mass density, $\widetilde{\sigma }$ the
electric charge density and $\widetilde{\nu }$ the kinematic viscosity.

In orthogonal curvilinear coordinates, denote by $\left( \widetilde{u},%
\widetilde{v},\widetilde{w}\right) $ the streamwise, the wall-normal and the
spanwise components of the physical velocity field $\widetilde{U}$. Consider
reference quantities, of the order of typical physical parameters, and
change to non-dimensional variables 
\begin{equation}
\begin{array}{llllll}
x=\frac{\widetilde{x}}{L_{r}}; & y=\frac{\widetilde{y}}{\delta _{r}}; & z=%
\frac{\widetilde{z}}{L_{r}}; & t=\widetilde{t}\frac{U_{r}}{L_{r}}; & u=\frac{%
\widetilde{u}}{U_{r}}; & v=\frac{\widetilde{v}L_{r}}{U_{r}\delta _{r}} \\ 
w=\frac{\widetilde{w}}{U_{r}}; & \rho _{m}=\frac{\widetilde{\rho }_{m}}{\rho
_{r}}; & p=\frac{\widetilde{p}}{\rho _{r}U_{r}^{2}}; & \nu =\frac{\widetilde{%
\nu }}{\nu _{r}}; & \sigma =\frac{\widetilde{\sigma }}{\sigma _{r}}; & E=%
\frac{\widetilde{E}}{E_{r}}
\end{array}
\label{2.3}
\end{equation}
Typical values for the reference quantities are

$U_{r}=100$ m s$^{-1}$, $L_{r}=1$ m, $\delta _{r}=10^{-3}$ m, $\rho _{r}=1.2$
Kg m$^{-3}$, $E_{r}=500$ V cm$^{-1}$, $\sigma _{r}=15$ $\mu $C cm$^{-3}$, $%
\nu _{r}=1.5\times 10^{-5}$ m$^{2}$ s$^{-1}$. Then R$_{L}=\frac{U_{r}L_{r}}{%
\nu _{r}}=6.66\times 10^{6}$ and $\frac{1}{R_{L}}$ and $\frac{\delta _{r}^{2}%
}{L_{r}^{2}}=10^{-6}$ are small quantities.

Expressing (\ref{2.1}) in the non-dimensional variables (\ref{2.3}),
assuming the product $k\delta $ (airfoil curvature $\times $ boundary layer
width) to be small and neglecting terms of order $R_{L}^{-1}$, $\frac{\delta
_{r}^{2}}{L_{r}^{2}}$ (and $\frac{\widetilde{U}}{c}$) one obtains 
\begin{equation}
\begin{array}{rll}
\frac{\partial u}{\partial t}+u\frac{\partial u}{\partial x}+v\frac{\partial
u}{\partial y}+w\frac{\partial u}{\partial z} & = & -\frac{1}{\rho _{m}}%
\frac{\partial p}{\partial x}+\nu \omega \frac{\partial ^{2}u}{\partial y^{2}%
}+F_{x} \\ 
0 & = & -\frac{1}{\rho _{m}}\frac{\partial p}{\partial y}+F_{y} \\ 
\frac{\partial w}{\partial t}+u\frac{\partial w}{\partial x}+v\frac{\partial
w}{\partial y}+w\frac{\partial w}{\partial z} & = & -\frac{1}{\rho _{m}}%
\frac{\partial p}{\partial z}+\nu \omega \frac{\partial ^{2}w}{\partial y^{2}%
}+F_{z}
\end{array}
\label{2.4}
\end{equation}
with $\gamma =\frac{L_{r}\sigma _{r}E_{r}}{U_{r}^{2}\rho _{r}}=62.5,\omega =%
\frac{L_{r}^{2}}{\delta _{r}^{2}R_{L}}$ $=0.15$ and, for the ionized fluid
case, $F_{x}=\frac{\gamma }{\rho _{m}}\sigma E_{x}$ ; $F_{y}=\frac{\delta
_{r}}{L_{r}}\gamma \sigma E_{y}$ ; $F_{z}=\frac{\gamma }{\rho _{m}}\sigma
E_{z}$

Split the variables into steady state ($\overline{u},...$) and fluctuating
components ($u^{^{\prime }},...$) 
\[
\begin{array}{lll}
u & = & \overline{u}+u^{^{\prime }} \\ 
v & = & \overline{v}+v^{^{\prime }} \\ 
w & = & \overline{w}+w^{^{\prime }} \\ 
p & = & \overline{p}+p^{^{\prime }}
\end{array}
\]
Experimentally a sequence roll-streak-burst is a localized event in some
small region of the boundary layer. Therefore it is reasonable to use, for
the discussion of the coherent structures in that region, a quasiparallel
assumption for the steady-state solution, namely 
\[
\overline{v}=\overline{w}=\frac{\partial \overline{u}}{\partial x}=\frac{%
\partial \overline{u}}{\partial z}=0 
\]
For example, for a scaling solution \cite{Vilela1} 
\[
\overline{u}=u_{e}\left( 1-\exp \left( -y\chi \right) \right) 
\]
with $\chi =\sqrt{\frac{\gamma \sigma _{0}\overline{E}_{x}}{u_{e}\omega \nu
\rho _{m}}}$\ and $\sigma =\sigma _{0}\left( 1-\frac{\overline{u}}{u_{e}}%
\right) $

By differentiation of Eqs.(\ref{2.4}) the pressure terms may be eliminated.
Then, keeping only the linear terms in the fluctuating fields one obtains 
\begin{equation}
\begin{array}{rcl}
\frac{\partial }{\partial t}\frac{\partial u^{^{\prime }}}{\partial y}+\frac{%
\partial }{\partial y}\left( \overline{u}\frac{\partial u^{^{\prime }}}{%
\partial x}+v^{^{\prime }}\frac{\partial \overline{u}}{\partial y}\right)
-\nu \omega \frac{\partial ^{3}u^{^{\prime }}}{\partial y^{3}} & = & \frac{%
\partial F_{x}^{^{\prime }}}{\partial y}-\frac{\partial F_{y}^{^{\prime }}}{%
\partial x} \\ 
\frac{\partial }{\partial t}\frac{\partial w^{^{\prime }}}{\partial y}+\frac{%
\partial }{\partial y}\left( \overline{u}\frac{\partial w^{^{\prime }}}{%
\partial x}\right) -\nu \omega \frac{\partial ^{3}w^{^{\prime }}}{\partial
y^{3}} & = & \frac{\partial F_{z}^{^{\prime }}}{\partial y}-\frac{\partial
F_{y}^{^{\prime }}}{\partial z}
\end{array}
\label{2.5}
\end{equation}
where $F_{i}^{^{\prime }}$ are the fluctuating components of an eventual
controlling force. Without active control $F_{i}^{^{\prime }}=0$.\medskip 

\textbf{Solutions of (\ref{2.5}) (with }$F^{^{\prime }}=0$\textbf{)\medskip }

Let 
\begin{equation}
\begin{array}{lll}
u^{^{\prime }} & = & \frac{1}{\xi }e^{\xi x}\cos (\beta z)U(y,t) \\ 
v^{^{\prime }} & = & e^{\xi x}\cos (\beta z)V(y,t) \\ 
w^{^{\prime }} & = & \frac{1}{\beta }e^{\xi x}\sin (\beta z)W(y,t)
\end{array}
\label{2.7}
\end{equation}
with 
\[
U(y,t)+\frac{dV(y,t)}{dy}+W(y,t)=0 
\]
and boundary conditions 
\begin{eqnarray*}
U(0,t) &=&U(\delta _{*},t)=V(0,t)=V(\delta _{*},t) \\
&=&W(0,t)=W(\delta _{*},t)=0
\end{eqnarray*}
$\delta _{*}$ being a wall-normal size for the coherent structure, of the
order of the boundary layer width. The qualitative nature of the solution
does not depend much on the form that one assumes for the steady-state
solution. In the following one considers an exponential profile.

\begin{equation}
\overline{u}=u_{e}\left( 1-\exp \left( -y\frac{k}{\delta _{*}}\right) \right)
\label{2.8}
\end{equation}
Similar results are obtained, for example, for a linear profile

\# \textbf{Zero-order solution} (to be used as an initial condition for the
full solution):

If $\overline{u}$ is replaced by its average value $\overline{u}_{m}$ in the
boundary layer 
\[
\overline{u}_{m}=u_{e}\left( 1-\frac{1}{k}+\frac{e^{-k}}{k}\right) 
\]
one would obtain 
\begin{equation}
\begin{array}{lll}
U(y,t) & = & c_{u}e^{\lambda t}\sin (\frac{2\pi }{\delta _{*}}y) \\ 
V(y,t) & = & -\frac{\delta _{*}}{2\pi }c_{v}e^{\lambda t}\left( \cos (\frac{%
2\pi }{\delta _{*}}y)-1\right) \\ 
W(y,t) & = & c_{w}e^{\lambda t}\sin (\frac{2\pi }{\delta _{*}}y)
\end{array}
\label{2.9}
\end{equation}
with $c_{u}+c_{v}+c_{w}=0$ and 
\begin{equation}
\lambda =-\xi \overline{u}_{m}-\nu \omega \left( \frac{2\pi }{\delta _{*}}%
\right) ^{2}  \label{2.10}
\end{equation}
This implies that for the solution to grow in time it must decrease in $x-$%
space, namely 
\begin{equation}
\xi \lesssim -\frac{\nu \omega }{\overline{u}_{m}}\left( \frac{2\pi }{\delta
_{*}}\right) ^{2}  \label{2.11}
\end{equation}
That is, the rolls-plus-streaks instability leads to a localized turbulence
spot and $x$-dependence is seen to be an essential ingredient to have a
sustained structure. The $zy$ and $zx$ structure of this averaged-profile
solution are shown in Fig.1a,b.

This approximate solution is now used as initial condition for the solutions
of Eqs.(\ref{2.5}). Reintroducing the $y$-dependence of $\overline{u}$,
using the continuity equation and (\ref{2.5}) and integrating once in $y$ 
\begin{equation}
\begin{array}{rrr}
\frac{\partial }{\partial t}\frac{dV}{dy}-\nu \omega \frac{d^{3}V}{dy^{3}}%
+\xi \overline{u}(y)\frac{dV}{dy}-\frac{d\overline{u}}{dy}\xi V & = & 
C_{V}(t) \\ 
\frac{\partial }{\partial t}W-\nu \omega \frac{\partial ^{2}W}{\partial y^{2}%
}+\xi \overline{u}(y)W & = & C_{W}(t)
\end{array}
\label{2.12}
\end{equation}
In Sect.4, in the discussion of active control by external body forces,
these equations will be handled in a direct way. Here however, a Fourier
mode expansion is used, which will be convenient for the phase randomization
discussion of Sect.3. Let 
\begin{equation}
\begin{array}{lll}
V(y,t) & = & \sum_{n}v_{n}(t)e^{in\frac{2\pi }{\delta _{*}}y} \\ 
W(y,t) & = & \sum_{n}w_{n}(t)e^{in\frac{2\pi }{\delta _{*}}y}
\end{array}
\label{2.13}
\end{equation}
with $\xi =-\gamma \frac{\nu \omega }{\overline{u}_{e}}\left( \frac{2\pi }{%
\delta _{*}}\right) ^{2}$ , $\chi =\frac{k}{\delta _{*}}$ and $\tau =\nu
\omega \left( \frac{2\pi }{\delta _{*}}\right) ^{2}t$ . Then 
\begin{equation}
\begin{array}{lll}
n\frac{dv_{n}}{d\tau } & = & \left( -n^{3}+\gamma n\frac{k-1+e^{-k}}{k}%
+i\gamma \frac{1-e^{-k}}{2\pi }\right) v_{n} \\ 
&  & +\gamma \sum\limits_{n^{^{\prime }}\neq n}\left( n^{^{\prime }}-i\frac{k%
}{2\pi }\right) v_{n^{^{\prime }}}g(n^{^{\prime }},n)+C_{v}(t)\delta _{n,0}
\\ 
\frac{dw_{n}}{d\tau } & = & \left( -n^{2}+\gamma \frac{k-1+e^{-k}}{k}\right)
w_{n}+\gamma \sum\limits_{n^{^{\prime }}\neq n}w_{n^{^{\prime
}}}g(n^{^{\prime }},n)+C_{w}(t)\delta _{n,0}
\end{array}
\label{2.14}
\end{equation}
and 
\begin{equation}
g(n^{^{\prime }},n)=\frac{1-e^{-k}}{i2\pi \left( n^{^{\prime }}-n\right) -k}
\label{2.15}
\end{equation}
For the $t=0$ initial condition use $\left( v_{0}=-\varepsilon ;v_{1}=v_{-1}=%
\frac{\varepsilon }{2};w_{1}=-w_{1}=\frac{\varepsilon }{2i}\right) $ and the
boundary conditions are $\sum_{n}v_{n}=0=\sum_{n}w_{n}$. The evolution of $%
v_{n}$ and $w_{n}$ (for $n\neq 0$) is obtained from (\ref{2.14}) and $v_{0}$
and $w_{0}$ are fixed by the boundary conditions. The allowed
time-dependence of $C_{V}$ and $C_{W}$ insures the consistency of the
process. The absolute value of the coefficients in Eq.(\ref{2.13}) obtained
after $t=5000\times 0.001$ time steps is shown in Fig.2 for the parameters ($%
n_{\max }=30$, $\gamma =5$, $k=2$). Fig.3 shows the $zy-$structure of this
solution. One sees that, starting from the symmetric initial condition in
Fig.1a, as time develops, the energy of the structure moves away from the
wall. Even Eq.(\ref{2.10}), although obtained for the averaged-profile
solution, already suggests a rate of growth increasing with the local
streamwise velocity. This behavior of the obtained solution matches the
experimental fact that unstable streaks lift away from the wall.

The solution, just studied, is sufficiently simple and well under control to
serve as a basis for the study of control methods. In the following section
the effect of random phases in this and other models is discussed and Sect.4
reports on the perspectives for active control by external forces.

\section{Random phases and energy transfer in the turbulence cascade}

Turbulence generation implies a transfer of energy between structures at
different length scales. Any mechanism that interferes with this energy
transfer may have a controlling effect on the rate of turbulence production.
A mechanism of this kind was put in evidence in Ref.\cite{Handler}. Using a
spectral code for a direct numerical simulation of a turbulent flow, it was
found that randomization of the phases of some of the Fourier modes have a
strong effect on turbulence and leads to drag reduction. Drag reduction was
also obtained on a recent experiment\cite{Sirovich2} of flow over a surface
with a random arrangement of protrusions. These effects have been
interpreted as originating from the destruction by the phase randomization
of the coherence of the large-scale turbulence producing structures.
Interfering with the formation of rolls and plane waves, random phases (or
random roughness) would delay turbulence production and bursting.

Another important effect that is associated to phase randomization is the
blocking of energy transfer to the small scales of turbulence. This is the
phenomenon that will be analyzed in this section, first in the context of
the shell model and then for the boundary layer equations.

\subsection{Random phases in the shell model}

The shell model is a toy model for the turbulence cascade. It may not be
directly relevant for the turbulence generating phenomena in the boundary
layer, nevertheless its study provides a good insight on the mechanisms of
energy transfer between length scales. Let $u_{n}$ be the Fourier modes of a
velocity field. Then, the Gledzer-Yamada-Ohkitani\cite{Gledzer} \cite
{Ohkitani} shell model is defined by the equation

\begin{eqnarray}
&&\left( \frac{d}{dt}+\nu k_{n}^{2}\right) u_{n}  \label{3.1} \\
&=&i\left( k_{n}u_{n+1}^{*}u_{n+2}^{*}-\frac{k_{n-1}}{2}%
u_{n-1}^{*}u_{n+1}^{*}-\frac{k_{n-2}}{2}u_{n-1}^{*}u_{n-2}^{*}\right)
+f\delta _{n,4}  \nonumber
\end{eqnarray}
The energy equation is 
\begin{equation}
\frac{d}{dt}\sum_{n=1}^{N}\left| u_{n}\right| ^{2}=-\nu
\sum_{k=1}^{N}k_{n}^{2}\left| u_{n}\right| ^{2}+\left(
f^{*}u_{4}+fu_{4}^{*}\right)   \label{3.2}
\end{equation}

To study the effect of random phases in the shell model, the numerical
scheme used is the same as in Ref.\cite{Pisarenko}, namely $k_{n}=k_{0}2^{n}$
\begin{equation}
u_{n}(t+dt)=e^{-\nu k_{n}^{2}dt}u_{n}(t)+\frac{1-e^{-\nu k_{n}^{2}dt}}{\nu
k_{n}^{2}}\left( \frac{3}{2}g_{n}(t)-\frac{1}{2}g(t-dt)\right)   \label{3.3}
\end{equation}
with 
\begin{equation}
g_{n}=ik_{n}\left( u_{n+1}^{*}u_{n+2}^{*}-\frac{1}{4}u_{n-1}^{*}u_{n+1}^{*}-%
\frac{1}{8}u_{n-1}^{*}u_{n-2}^{*}\right) +f\delta _{n,4}  \label{3.4}
\end{equation}
$k_{0}=2^{-4}$, $\nu =10^{-7}$, $dt=10^{-4}$, $f=0.005\left( 1+i\right) $

The system is driven at mode 4 and the other modes are energized by a
cascade of energy transfer. The simulations start from a point in the
attractor, the mode energy $\overline{u_{n}^{2}}$ being the average over $%
10^{4}$ time steps. The average mode energy in the attractor is shown in
Fig.4. It scales with a power law 
\[
\overline{u_{n}^{2}}\thicksim k_{n}^{-\frac{2}{3}} 
\]
in the inertial region.

Now, starting from a point in the attractor one lets the system evolve for a
time $\Delta t=130$ but, at each step, the phase of one or several of the
modes is randomized. The results are shown in Figs.5a-c. For Fig.5a
different random phases are given to the modes 10, 11 and 12. For Fig.5b
only the phase of mode 11 is randomized. For Fig.5c the same random phase is
given to the modes 10, 11 and 12. In all cases one observes a strong
blocking effect in the energy flow to the small scales, stronger when
several phases are randomized but also quite large when only the phase of
mode 11 is randomized. By contrast if, instead of random phases, one applies
a small random excitation to an intermediate mode almost no effect is
observed.

In this model the blocking of energy transfer by random phases has a simple
interpretation. The energy flow through mode $M$ is 
\begin{eqnarray}
&&\frac{d}{dt}\sum_{n=M+1}^{N}\left| u_{n}\right| ^{2}+2\nu
\sum_{k=1}^{N}k_{n}^{2}\left| u_{n}\right| ^{2}  \label{3.5} \\
&=&-2\textnormal{Im}\left( u_{M}u_{M+1}\left( k_{M}u_{M+2}+\frac{k_{M-1}}{2}%
u_{M-1}\right) \right) =2\pi _{M}  \nonumber
\end{eqnarray}

If $u_{M}$ is multiplied by a random phase 
\[
u_{M}\rightarrow u_{M}e^{i\theta (t)} 
\]
not correlated with $u_{M-1}$, $u_{M+2}$ and $u_{M+1}$, then 
\[
\left\langle \pi _{M}\right\rangle =0 
\]
and there is no energy flow through mode $M$. One also sees, from Eq.(\ref
{3.5}), that the possibility of blocking the energy transfer to the small
scales depends critically on the local nature of the energy flow. Namely,
the fact that only a few neighboring modes are involved in the energy
transfer. That means that any random phase, on any mode whatsoever, will
have a blocking effect. As it will be seen later, the situation in the
boundary layer equations is quite different.

\subsection{Random phases in the boundary layer equations}

For the boundary layer case one evolves Eqs.(\ref{2.14}), starting from the
same initial conditions and for the same number of time steps as in Sect.2.
Fig.6 shows the result when the phase of the $v_{1}$ component is randomized
at each step. In this case one obtains a sizable suppression of the energy
transfer to the higher modes. However, for the simulation parameters that
were used ($\gamma =5,k=2$), randomizing the phase of mode 1 is the only
case where any appreciable effect is observed. Randomizing the phases of the
higher modes does not seem to block the energy transfer.

This phenomenon may be understood from the Eqs.(\ref{2.14}). First, and
unlike the shell model, the nature of the transfer function $g(n^{^{\prime
}},n)$ implies that there are long-range effects, each mode receiving
contributions from all the others. On the other hand one sees from the
diagonal terms in Eqs.(\ref{2.14}) that, for each set of parameters, there
is a number of intrinsically growing modes up to some $n_{\max }$ , with the
other modes receiving energy from these unstable modes by the transfer
function $g(n^{^{\prime }},n)$. It so happens that, for the simulation
parameters that were used, mode 1 is the unstable mode and only if one acts
on this mode does one obtain some effect.

In view of these two phenomena (the long-range interaction and the stable
versus unstable mode effect) the method of random phases will be effective,
in the control of the structures, only if it is carefully designed to act on
all the unstable modes. It must be designed in a case-by-case basis and
might become ineffective when the kinematical parameters change.

\section{Active control. Prospects}

Here one analyses the effect of external forces (the $F^{^{\prime }}$ in Eq.(%
\ref{2.5})) on the growth and eventual stabilization of the coherent
structures in the boundary layer equations. The aim is to find out whether
an active control scheme may be developed using this type of forces. If
effective control is to be achieved, it is the dominant dynamical effects
that have to be addressed. Therefore it makes sense to use a setting as
simple as possible to be able to isolate and interpret the most basic
effects. In particular, one may assume that outside the boundary layer
region there is a uniform constant pressure without fluctuations. Then, if $%
F_{y}^{^{\prime }}=0$, the second equation in (\ref{2.4}) implies $\frac{%
\partial p}{\partial y}=0$ and therefore, from uniformity, also $\frac{%
\partial p}{\partial x}=\frac{\partial p}{\partial z}=0$. In this case the
pressure terms may be dropped.

Using now the information obtained in Sect.2, namely the fact that growing
coherent structures must be localized in the $x-$variable, one writes 
\begin{equation}
\begin{array}{lll}
u^{^{\prime }}(x,y,z,t) & = & e^{-\xi x}\frac{\partial }{\partial z}%
g(z)U(y,t) \\ 
v^{^{\prime }}(x,y,z,t) & = & e^{-\xi x}\frac{\partial }{\partial z}%
g(z)V(y,t) \\ 
w^{^{\prime }}(x,y,z,t) & = & e^{-\xi x}g(z)W(y,t)
\end{array}
\label{4.1}
\end{equation}
and the linearized fluctuation equations become 
\begin{equation}
\begin{array}{rll}
\frac{\partial U}{\partial t}-\xi \overline{u}(y)U+\frac{d\overline{u}(y)}{dy%
}V-\nu \omega \frac{\partial ^{2}U}{\partial y^{2}} & = & F_{x}^{^{\prime }}
\\ 
\frac{\partial W}{\partial t}-\xi \overline{u}(y)W-\nu \omega \frac{\partial
^{2}W}{\partial y^{2}} & = & F_{z}^{^{\prime }} \\ 
-\xi U+\frac{\partial V}{\partial y}+W & = & 0
\end{array}
\label{4.2}
\end{equation}
Notice that the linearized equations do not fix the $z-$dependence.

As before, consider an exponential profile for the steady-state flow 
\[
\overline{u}(y)=u_{e}\left( 1-e^{-\alpha y}\right) 
\]
Eqs.(\ref{4.2}) are simple enough to even allow explicit analytical
solutions. For example the second equation in (\ref{4.2}) reduces after a
change of variable to the modified Bessel equation. However, better insight
is obtained by displaying its numerical solution, starting from different
sets of initial conditions. For example, starting from an initial condition $%
U(y,t)\thicksim \sin (2\pi y)$, $V(y,t)\thicksim \left( \cos (2\pi
y)-1\right) $ and $W\thicksim \sin (2\pi y)$ (Fig.7a), as in Sect.2, one
finds once more that time evolution lifts this structure away from the wall
(Fig.7b).

To control this structure would mean to damp it or at least to avoid its
instability, growth and lifting away from the wall. Instability of the
structure generates small scale turbulence. On the other hand when the
structure lifts away from the wall one sees a clear growth of the $V-$%
component which is the mathematical equivalent of an ejection or a sweep.
This exchange of boundary layer fluid with the external fast moving fluid
increases the skin friction drag.

Mathematically it would be very simple to invent forcing terms $%
F_{x}^{^{\prime }}$ and $F_{z}^{^{\prime }}$ which stabilize or damp the
structures. For example, forces proportional to the second derivative $\frac{%
\partial ^{2}}{\partial y^{2}}$ would be equivalent to a change of the
viscosity coefficient . However it is not realistic to assume that forces
with the appropriate $y-$dependence may be obtained, nor it is in general
possible to detect the corresponding $y-$variation of the fluctuating flow
components. At most one may apply a force that is essentially constant
throughout the boundary layer and the signal to be detected might be the
value of the derivative at $y=0$ of the velocity fluctuations. In Fig.7c the
result of a control experiment of this type is displayed. At each time step
the derivatives $\frac{\partial U}{\partial y}|_{y=0}$ and $\frac{\partial W%
}{\partial y}|_{y=0}$ are detected and a force is applied to the fluid near
the wall that is proportional to the difference between this derivative and
the corresponding derivative of the initial condition. That is, the control
tends to preserve the derivative at the wall. The range of the force that is
applied is the same as the range of the boundary condition (a unit
interval). One sees that the control maintains the circulation in the wall
region but does not avoid the lifting of the structure. Actually a linear
control, as used, has an overshooting effect because, as the structure lifts
away from the wall, the derivative naturally decreases and the control, that
at first counters the sweep in the $V-$component, later causes an ejection.

Instead of trying to stabilize a formed roll-streak structure one might
instead apply a force to kill the derivatives at the wall. In Figs.8a,b one
shows the results of such an experiment. Starting from a small random
non-zero initial condition in the interval $y\in [0,1]$ , a lifted structure
develops (Fig.8a) which is qualitatively similar to those obtained from
roll-streak initial conditions. Applying a controlling force proportional to
the derivatives at $y=0$ , one obtains the result shown in Fig.8b. One sees
that away from the wall the lifted structure is similar in the controlled
and non-controlled cases. The difference is the vanishing of the derivatives
at the wall in the controlled case. The conclusion is that, with detection
of the wall derivatives and an applied force in the boundary layer region,
the effect over the development, growth and eventual break-up of the
coherent structures is minimal. The only non-negligible effect that can be
achieved is that, by suppressing the wall derivatives, all this activity is
kept away from a small near-wall region. In practical terms, the conclusion
is that not much effect is be expected in the turbulence production rate.
However, by keeping a non-active small layer at the wall, a reduction in the
skin friction drag might be obtained. The effect is similar to one of the
interpretations of a riblet structure, namely that of keeping a region of
non-active fluid inside the grooves. The difference is that here no penalty
would be paid for an increased contact surface.

\section{Conclusions}

Isolating the dominating terms in the near-wall flow equations, some
understanding of the turbulence-generating structures is obtained. Also, the
simplified equations make a clear distinction between what is important and
what it is not, for the flow control action.

Phase randomness (introduced by surface effects, for example) which may be
very striking in particular models and for some kinematical regions, seems
to lack general validity and to require fine tuning of the phases that are
to be randomized.

Active control by external body forces, reacting to the measurement of the
surface derivatives, have only a limited effect on the development, growth
and break-up of the coherent structures. Therefore the turbulence production
rate is not significantly changed. However, by controlling a small layer of
non-active fluid near the wall, a reduction in the skin friction drag may be
obtained.

The type of externally controlled body forces that have been proposed in the
past\cite{Vilela1} \cite{Vilela2} may be obtained by injecting ionized fluid
(the same or another easier to ionize fluid) in the boundary layer and
acting on it by electromagnetic forces. In this setting, static fields
already have an effect on the flow profile\cite{Vilela1}. However to act on
the turbulence-generating structures active control is required. Special
attention has to be paid to the fact that, although the control acts on the
bulk of the boundary layer, the detection of the signals is only realistic
at the wall. Surface stress being directly related to wall-normal velocity
derivatives, this seems to be the most appropriate parameter on which to
base active control. Detection of field fluctuations, in an ionized fluid,
would be an interesting possibility, but it seems a more remote possibility
from a practical point of view.\medskip 

\textbf{Figure captions\medskip }

Fig.1a,b - $zy$ (a) and $zx$ (b) structure of the averaged-profile solution

Fig.2 - Fourier coefficients of the solution of Eq.(\ref{2.13}) after $5000$
time steps ($\Delta t=\times 0.001$)

Fig.3 - $zy-$structure of the solution corresponding to the Fourier
coefficients in Fig.2

Fig.4 - Average mode energy in the shell model attractor

Fig.5a-c - Average mode energy in the shell model with random phases in (a)
modes 10, 11 and 12, (b) mode 11 and (c) the same random phase in the modes
10, 11 and 12

Fig.6 - Fourier coefficients of the solution of Eq.(\ref{2.13}) when the
phase of the $v_{1}$ component is randomized

Fig.7a-c - Initial condition (a) and time evolved solution of Eqs.(\ref{4.2}%
) without (b) and with differential derivative active control (c)

Fig.8a,b - Time evolved solution of Eqs.(\ref{4.2}), starting from a random
initial condition without (a) and with derivative active control (b)

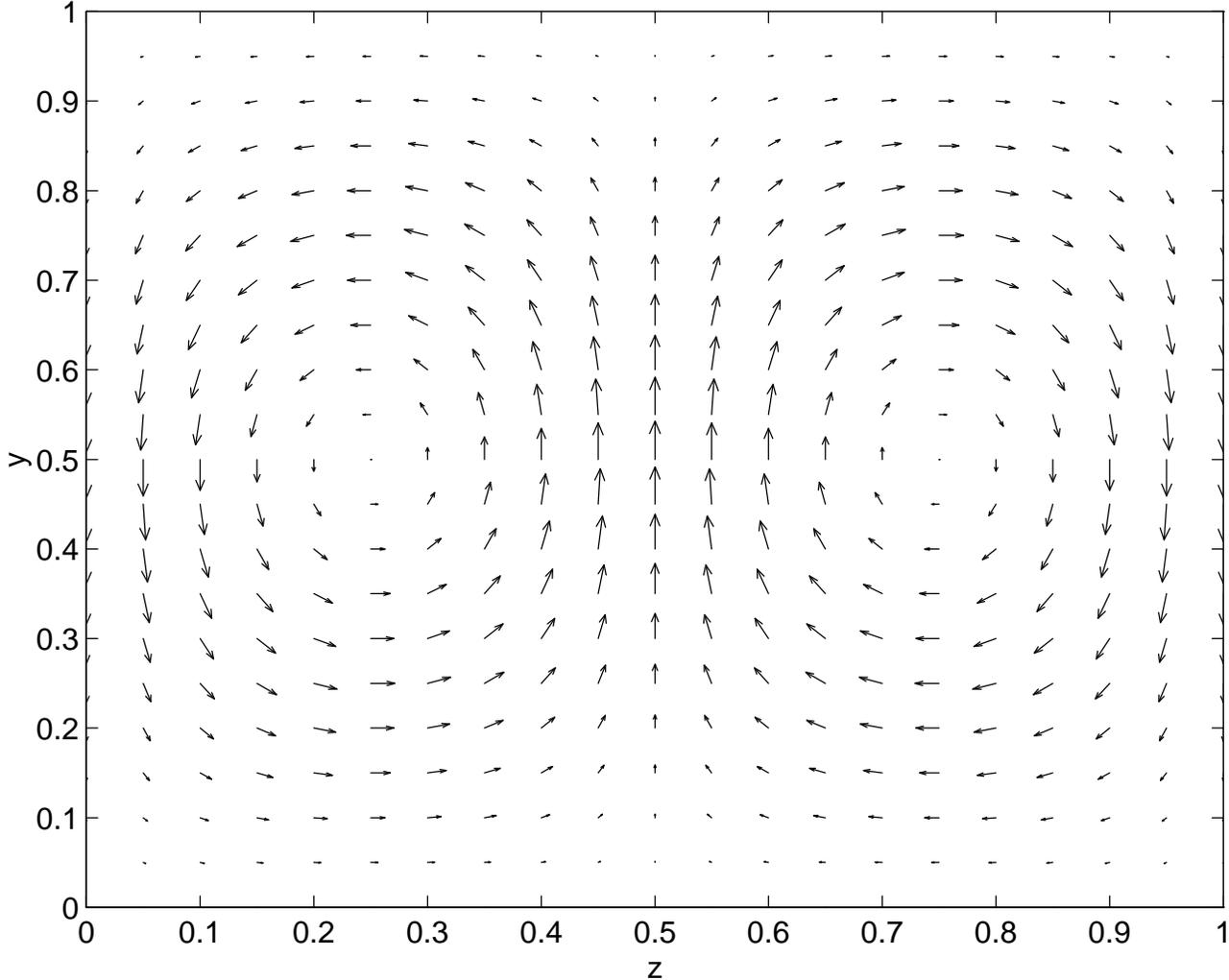
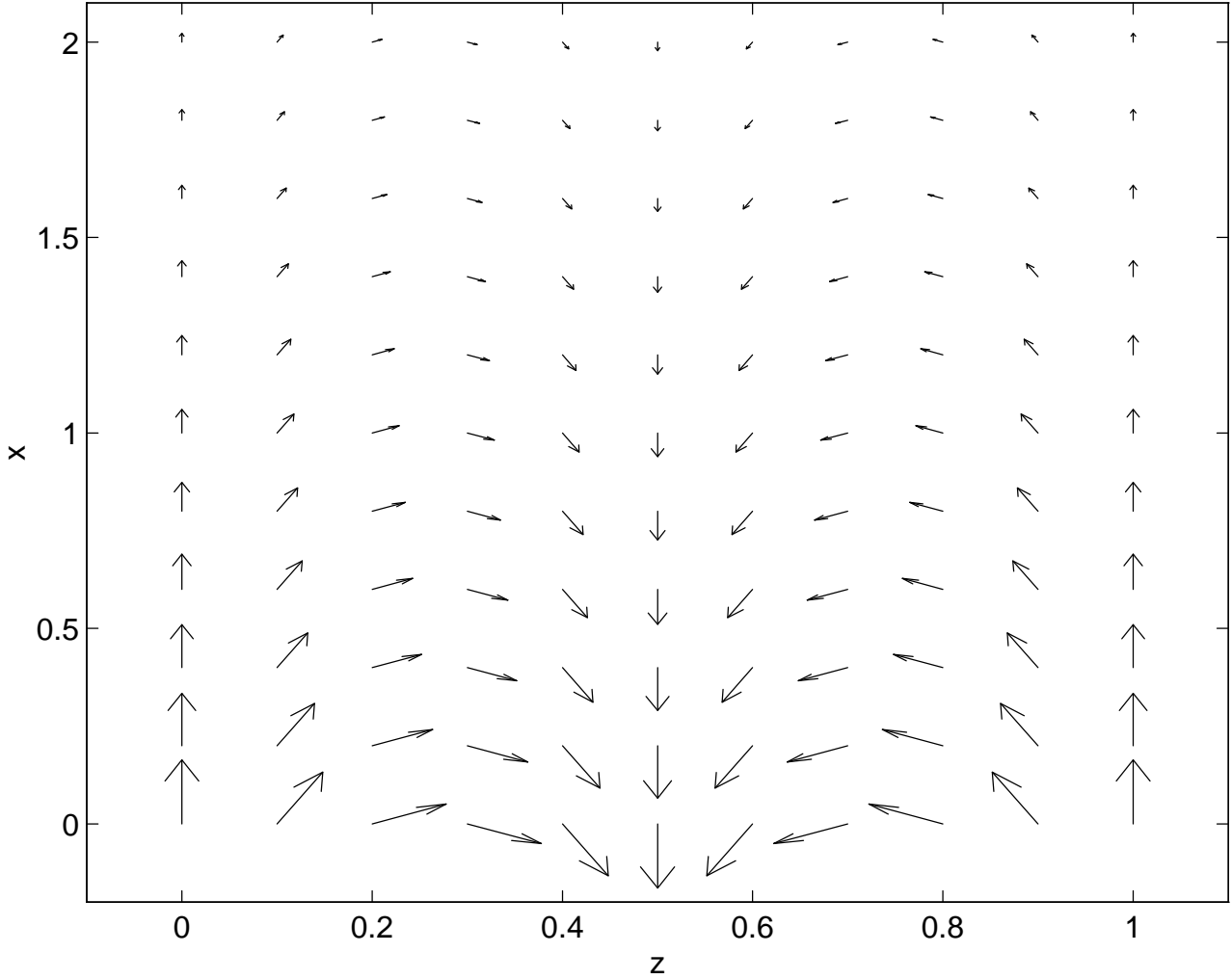
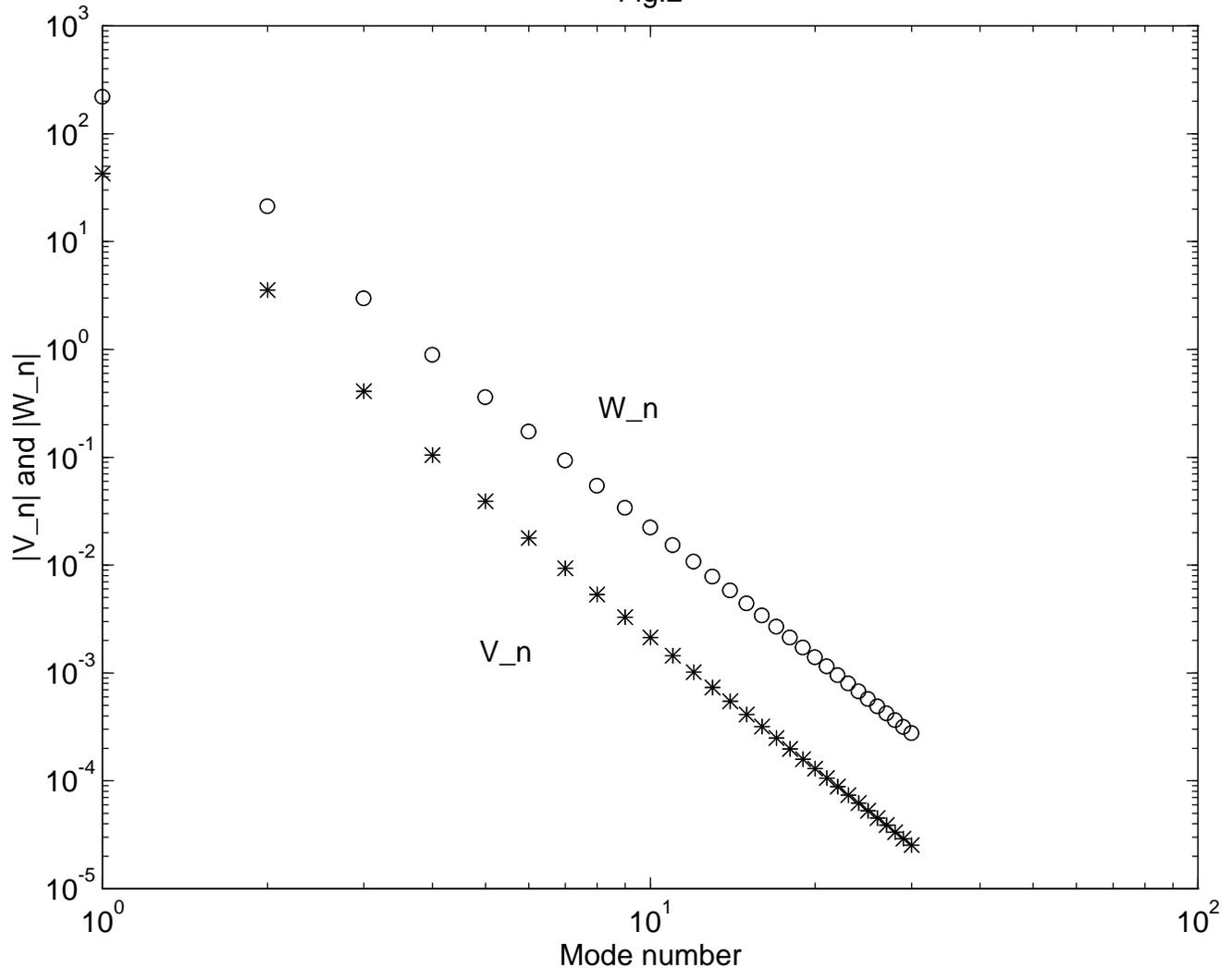
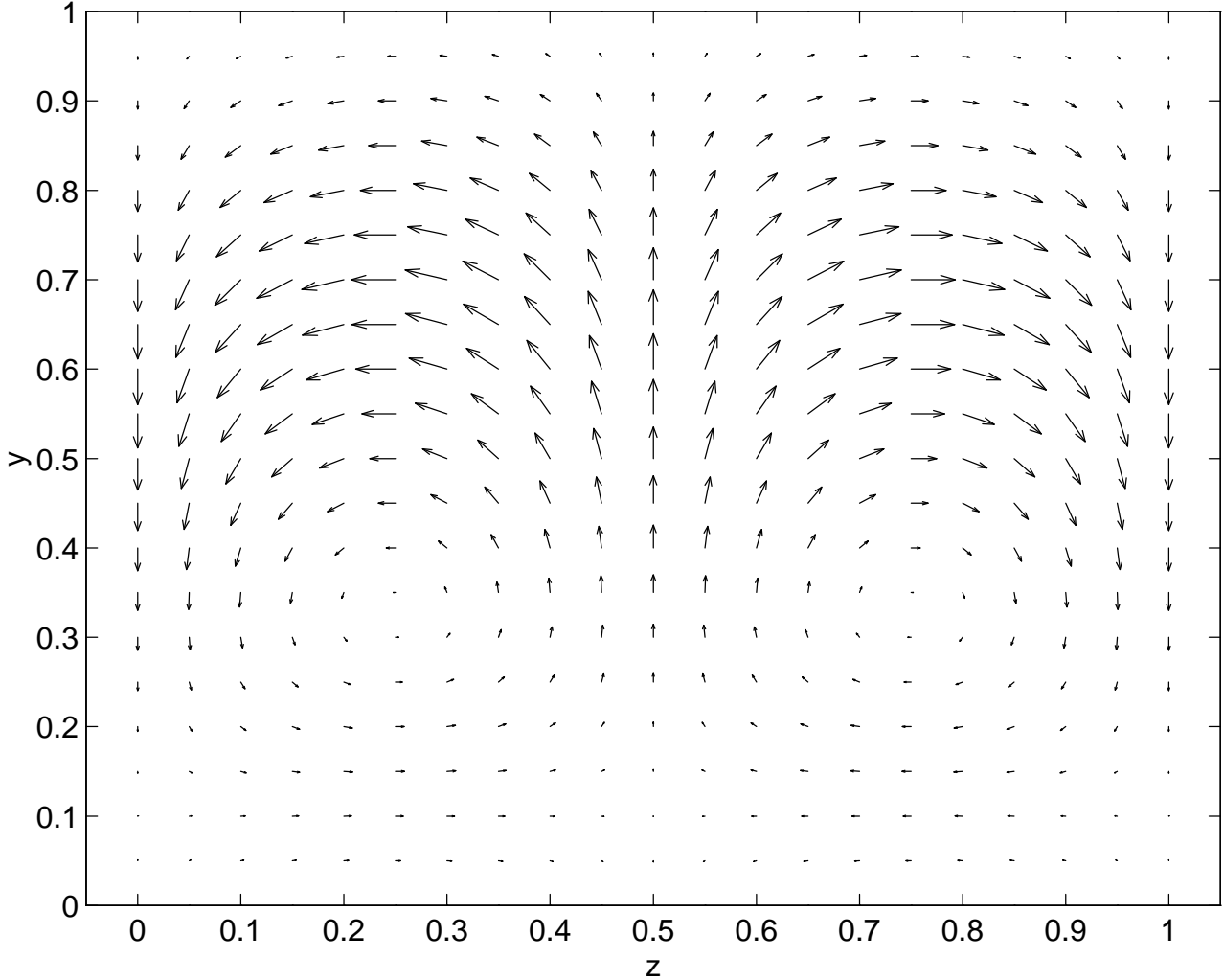
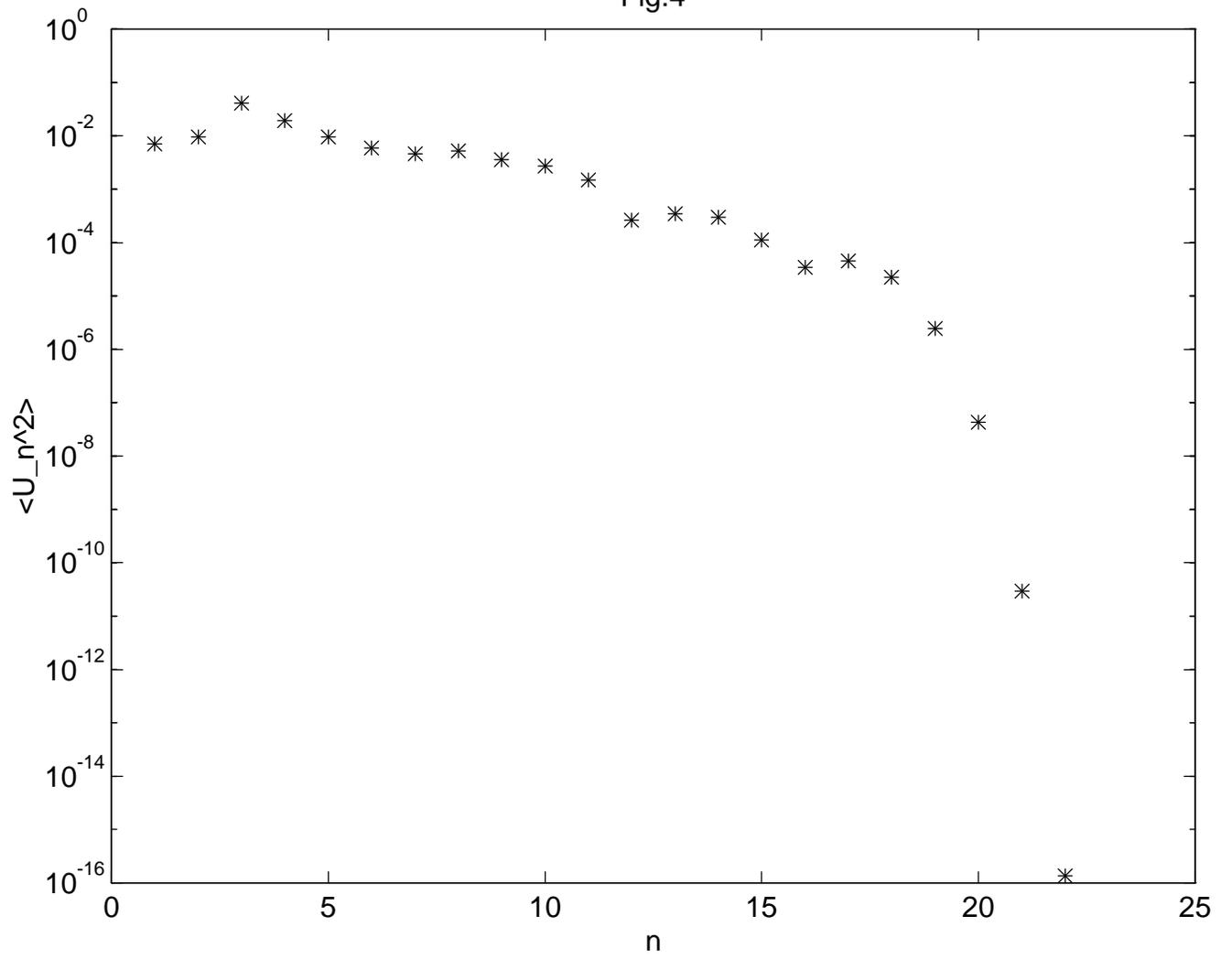
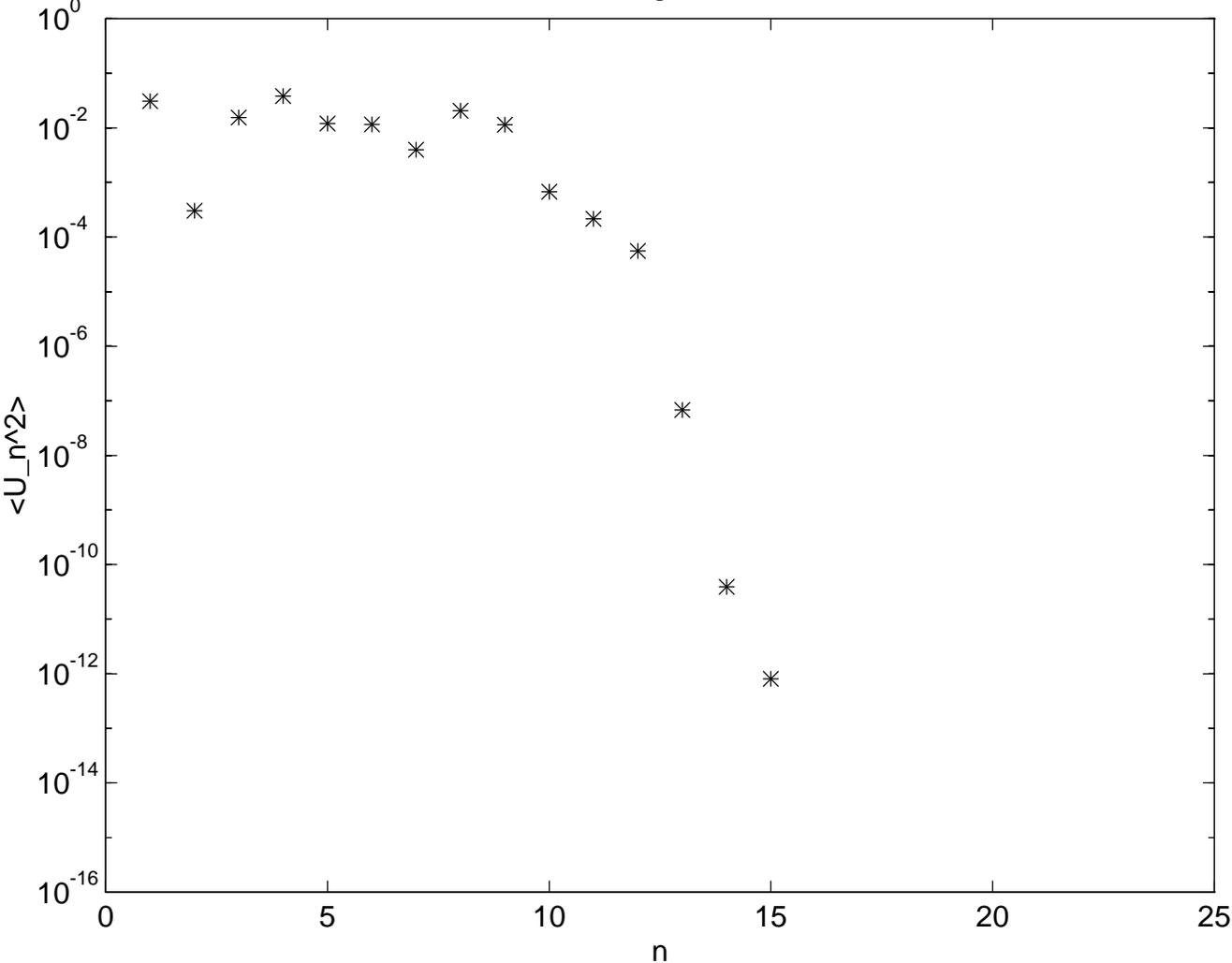
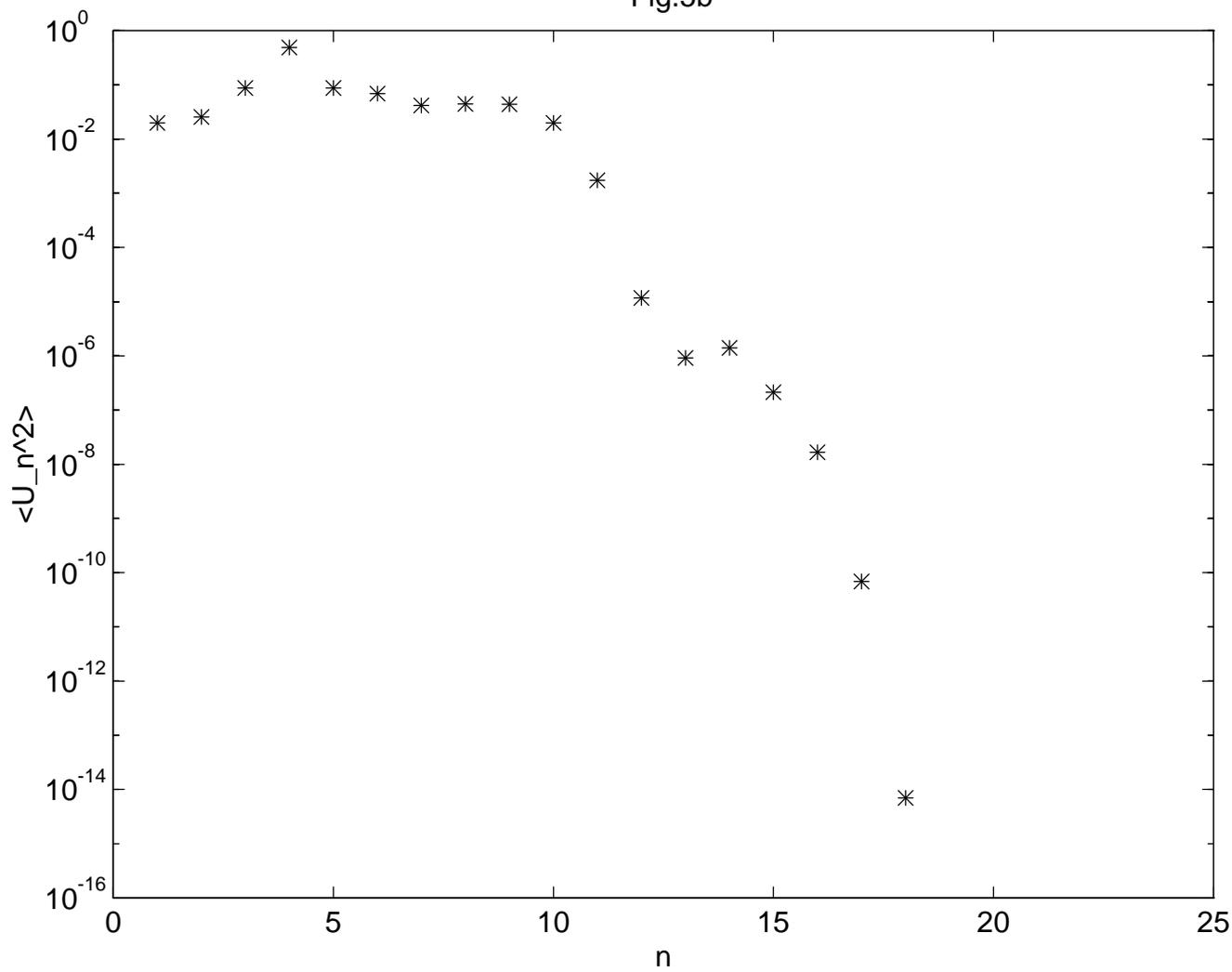
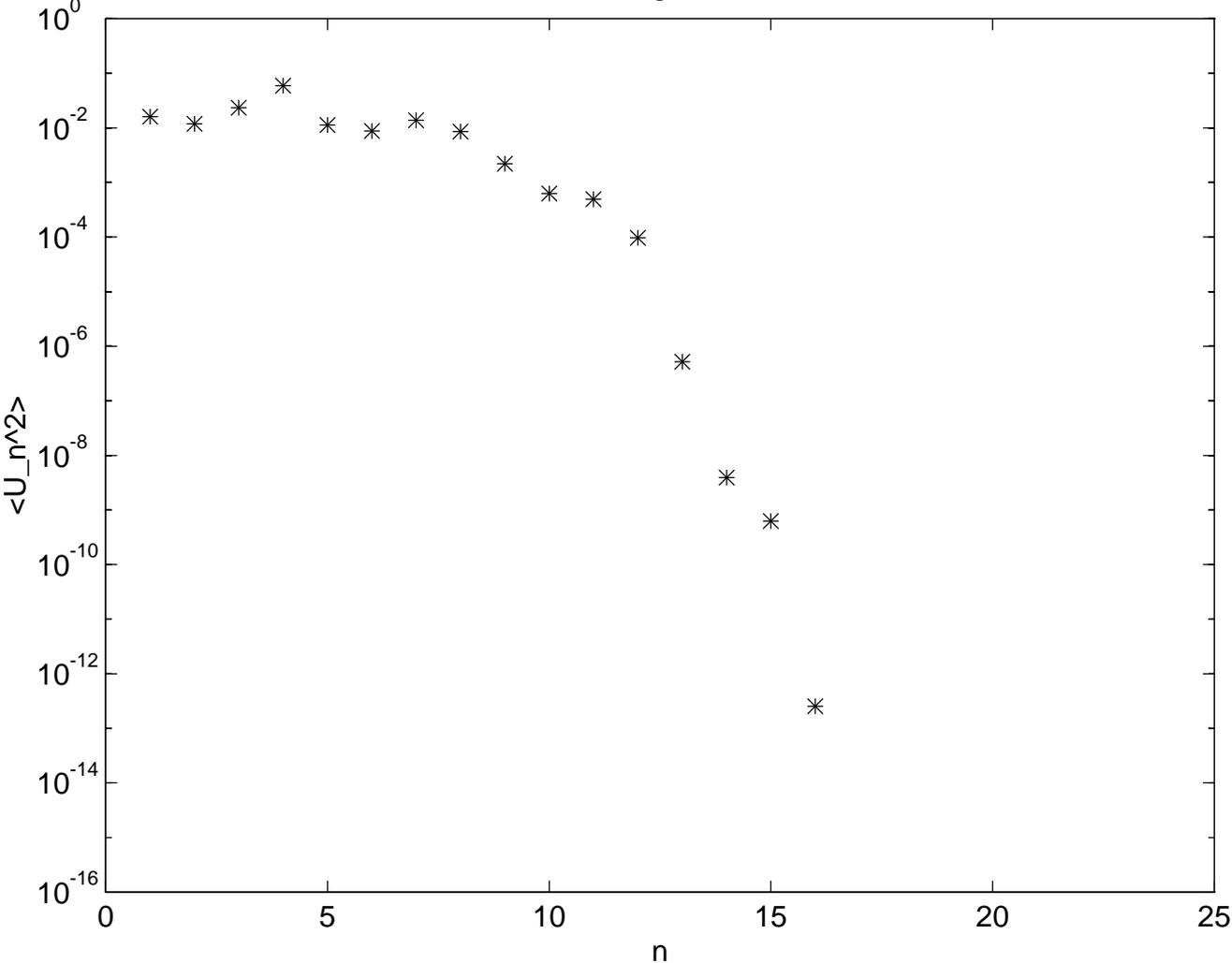
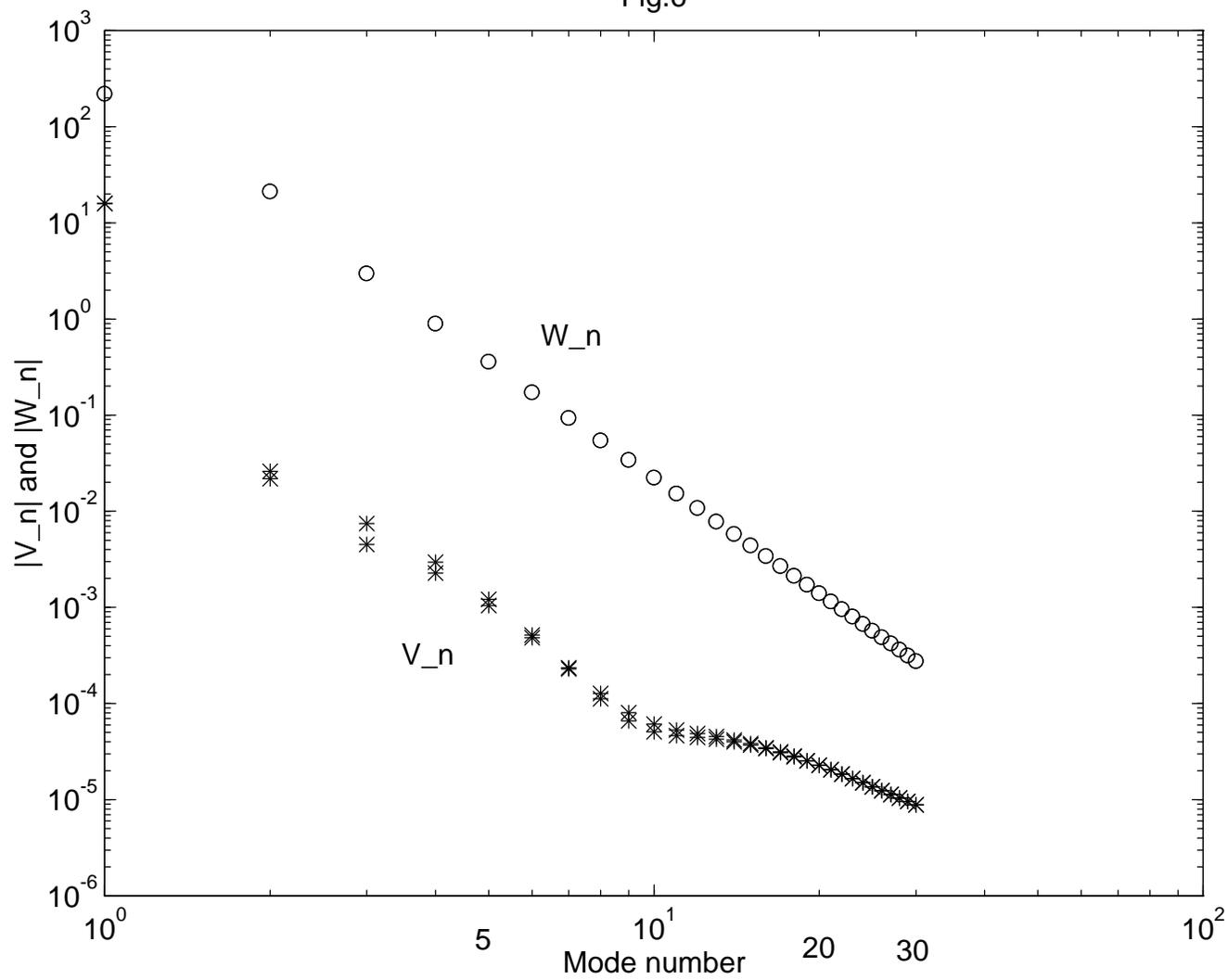
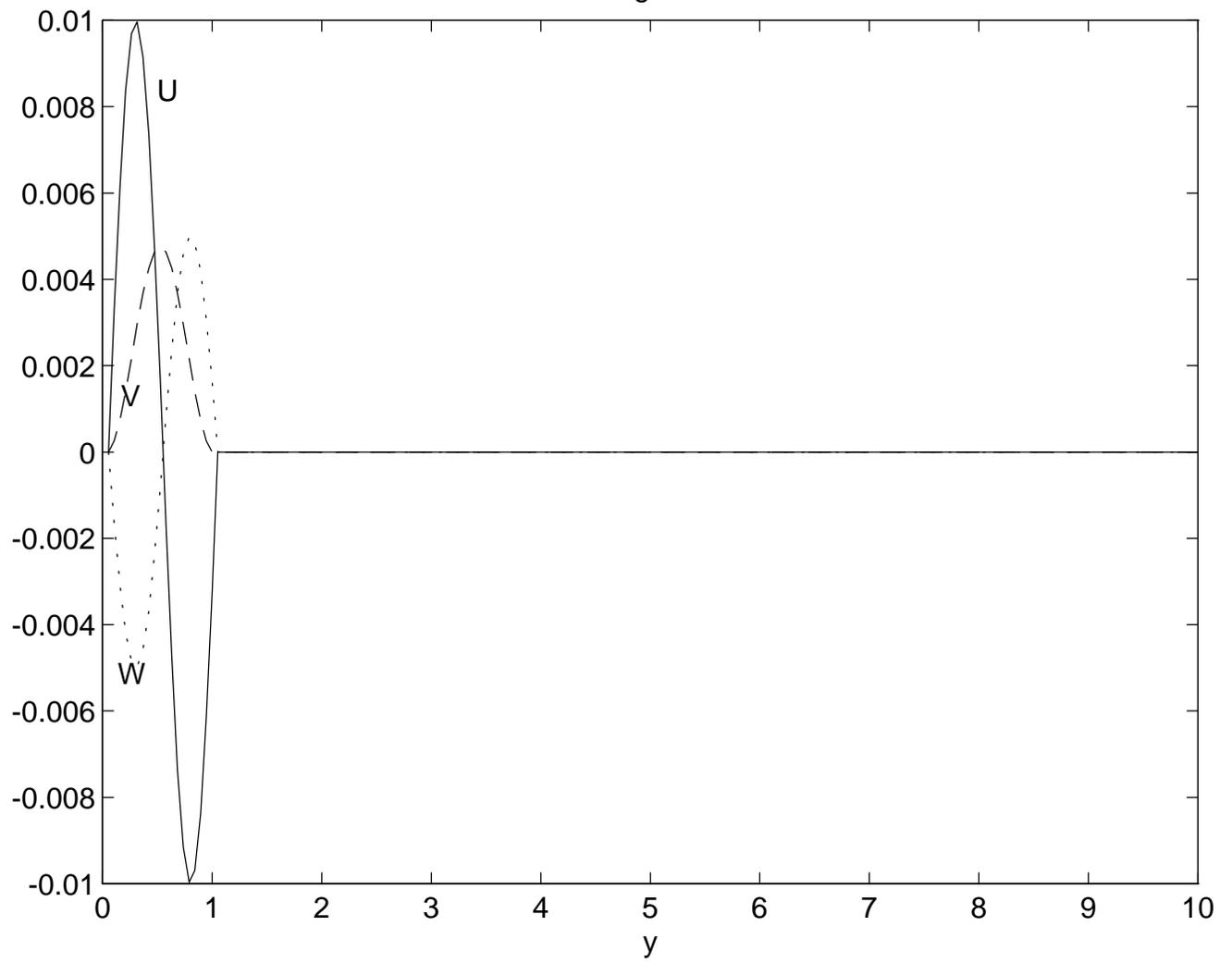
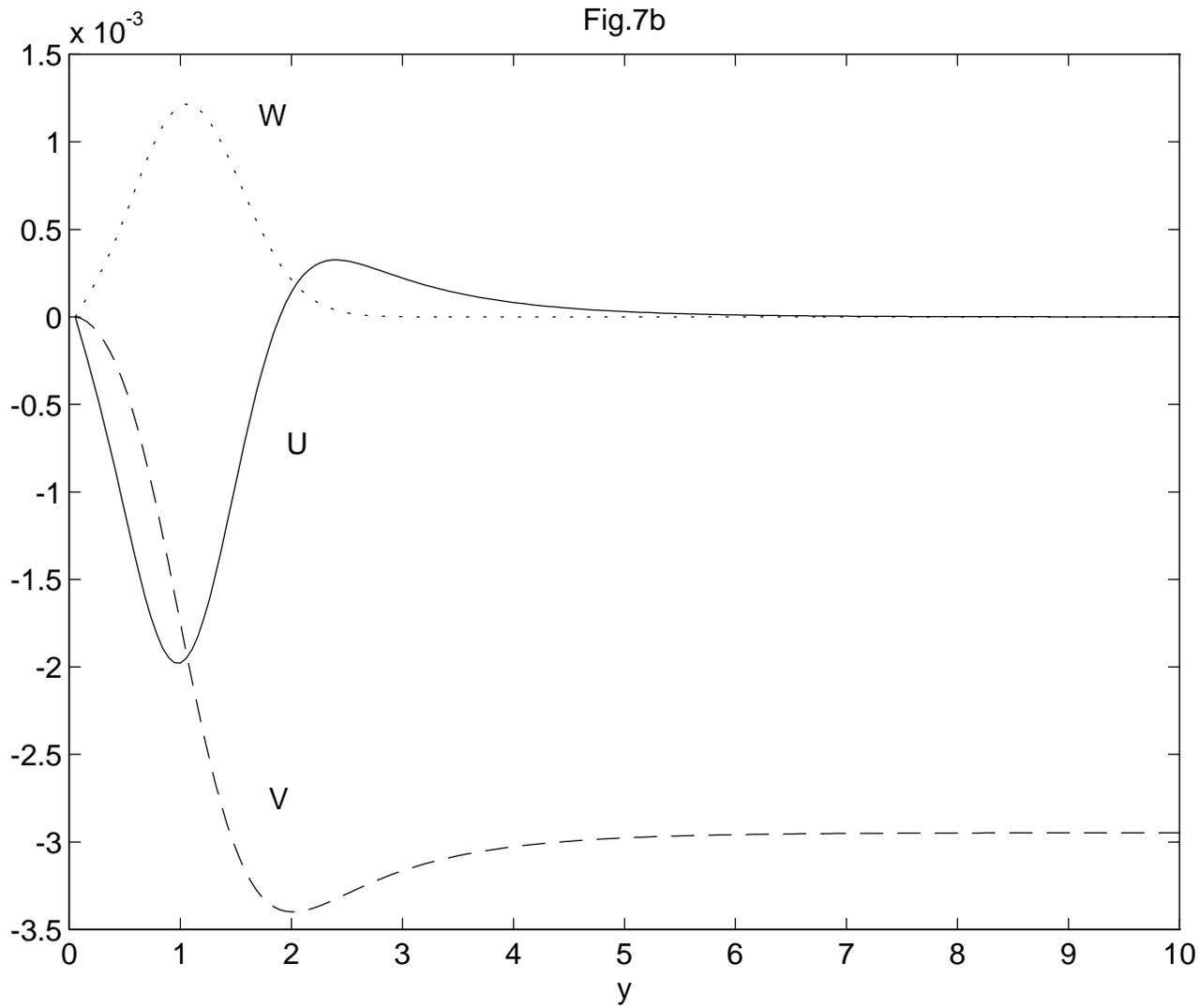
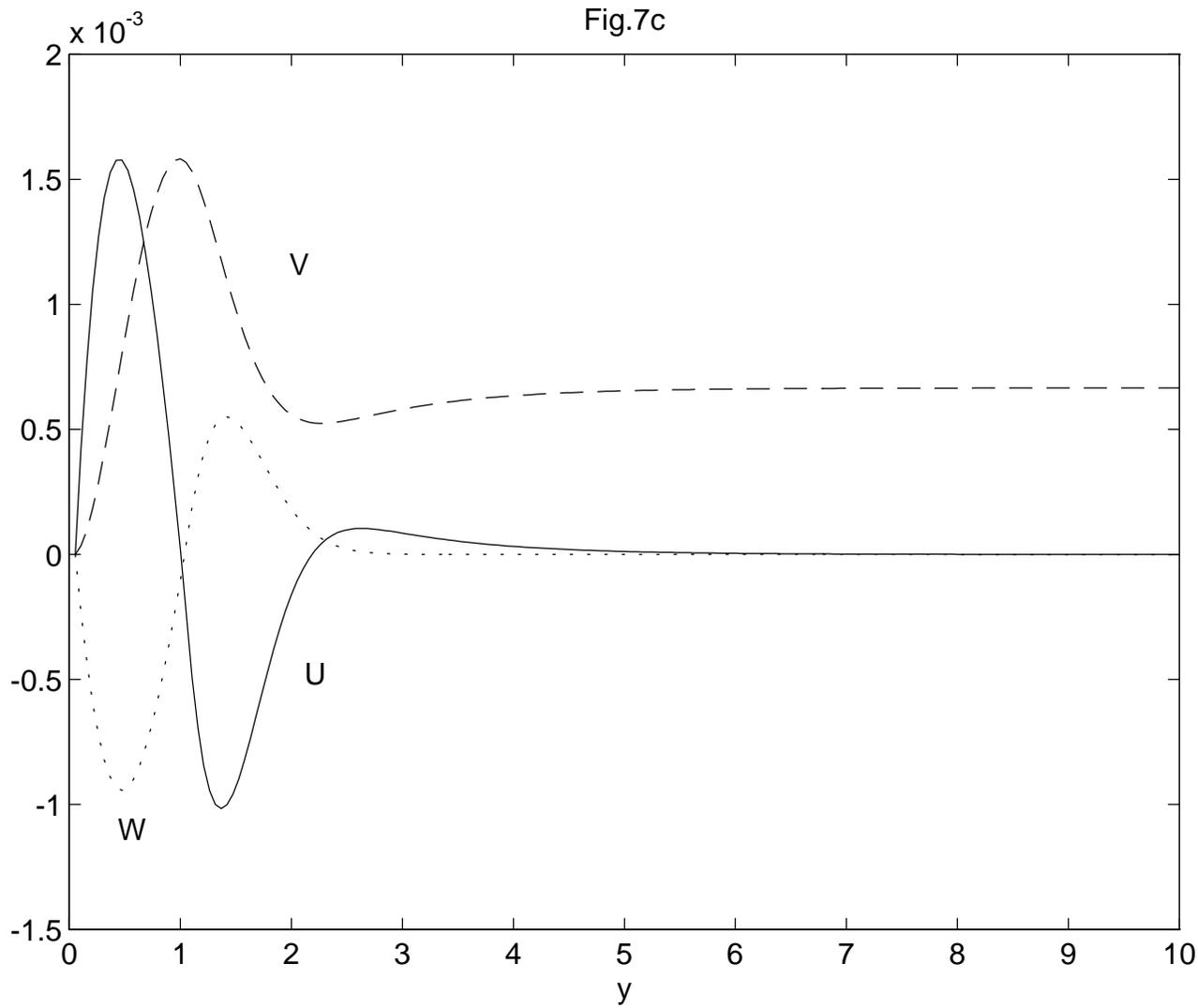
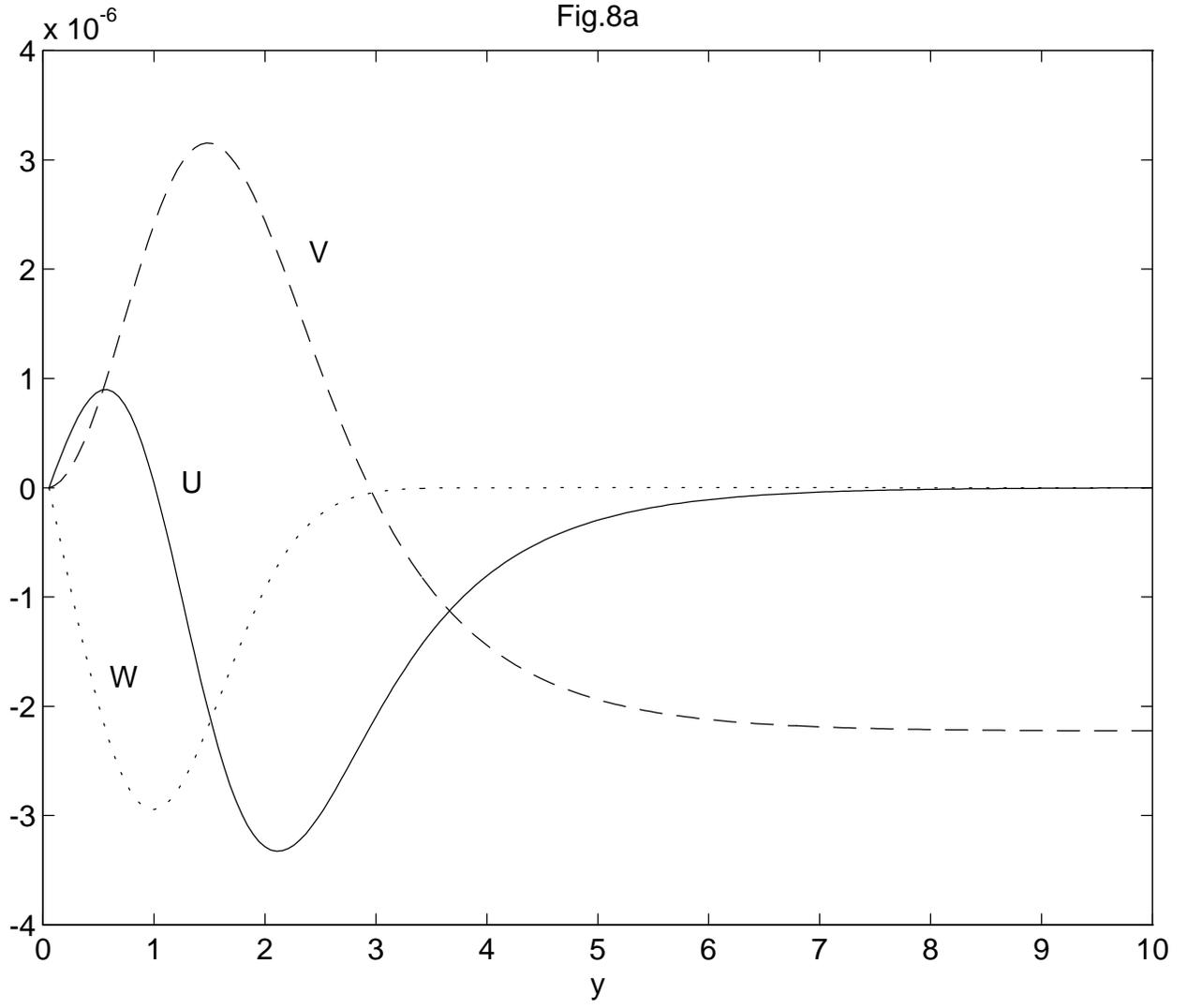
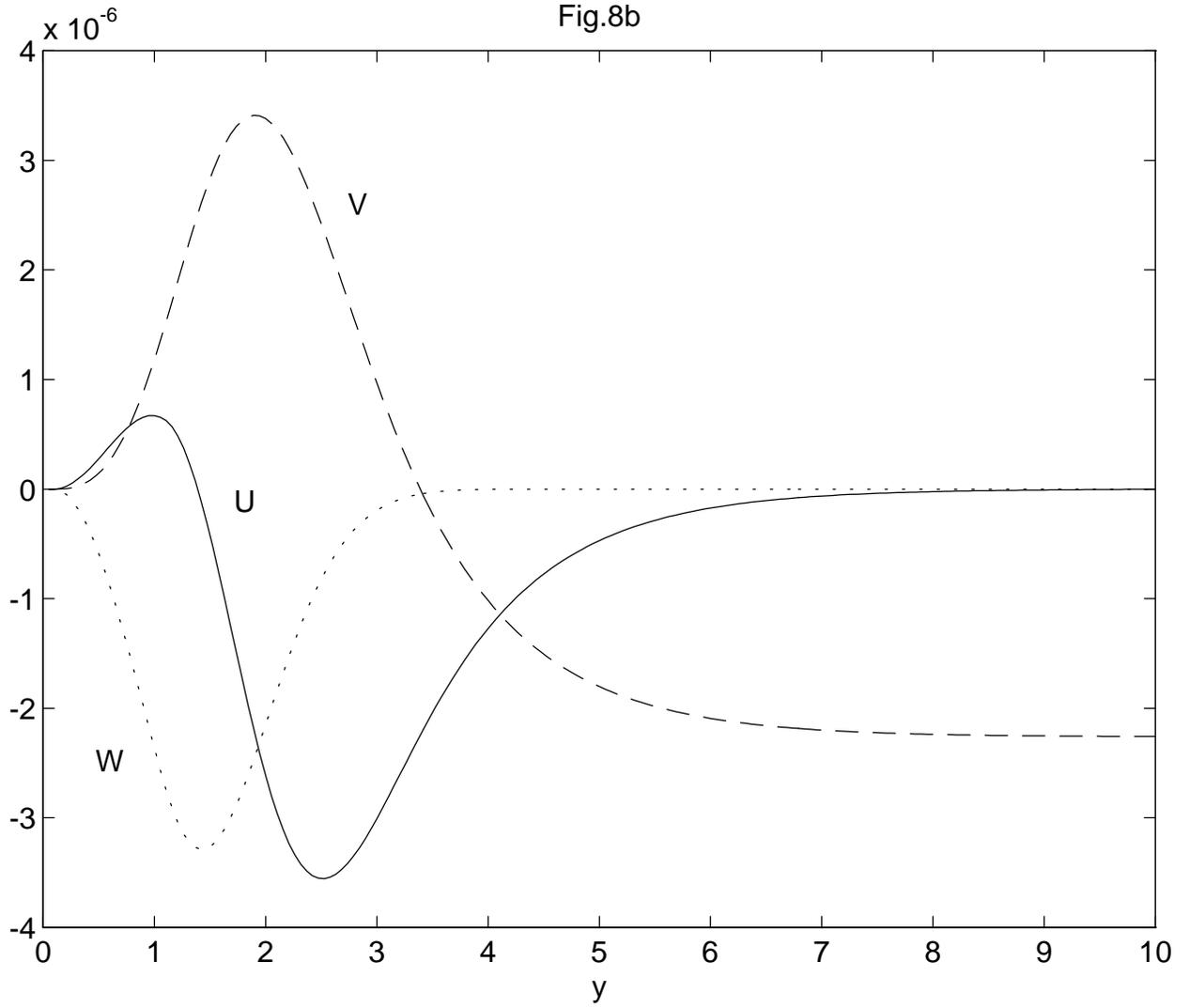
\end{document}